\documentclass[prb,showpacs,twocolumn,superscriptaddress]{revtex4-1}
\usepackage{graphicx,hyphenat}
\usepackage{color}

\def\z{\rule{0pt}{1.5ex}0}
\def\b{\rule{0pt}{3ex}Bethe}
\def\q{\rule{0pt}{3ex}QMC}

\renewcommand{\vec}[1]{\mathbf{#1}}

\begin{document}
	
\title{Magnetic structure and Dzyaloshinskii-Moriya interaction in the $S = 1/2$ helical-honeycomb antiferromagnet $\alpha$-Cu$_{2}$V$_{2}$O$_{7}$}
	
\author{G. Gitgeatpong}
\affiliation{Department of Physics, Faculty of Science, Mahidol University, Bangkok 10400, Thailand}
\affiliation{ThEP, Commission of Higher Education, Bangkok, 10400, Thailand}

\author{Y. Zhao}
\affiliation{Department of Materials Science and Engineering, University of Maryland, College Park, Maryland 20742, USA}
\affiliation{NIST Center for Neutron Research, National Institute of Standards and Technology, Gaithersburg, Maryland 20899, USA}
	
\author{M. Avdeev}
\affiliation{Australian Nuclear Science and Technology Organisation, ANSTO, Locked Bag 2001, Kirrawee DC, NSW, Australia}
	
\author{R. O. Piltz}
\affiliation{Australian Nuclear Science and Technology Organisation, ANSTO, Locked Bag 2001, Kirrawee DC, NSW, Australia}
	
\author{T. J. Sato}
\affiliation{IMRAM, Tohoku University, Sendai, Miyagi 980-8577, Japan}
	
\author{K. Matan}
\email[]{kittiwit.mat@mahidol.ac.th}
\affiliation{Department of Physics, Faculty of Science, Mahidol University, Bangkok 10400, Thailand}
\affiliation{ThEP, Commission of Higher Education, Bangkok, 10400, Thailand}
	
\date{\today}
	
\begin{abstract}
Magnetic properties of the $S = 1/2$ antiferromagnet $\alpha$-Cu$_{2}$V$_{2}$O$_{7}$ have been studied using magnetization, Quantum Monte Carlo (QMC) simulations, and neutron diffraction. Magnetic susceptibility shows a broad peak at $\sim50$~K followed by an abrupt increase indicative of a phase transition to a magnetically ordered state at $T_{N}$ = 33.4(1) K. Above $T_N$, a fit to the Curie-Weiss law gives a Curie-Weiss temperature of $\Theta=-73(1)$~K suggesting the dominant antiferromagnetic coupling. The result of the QMC calculations on the helical-honeycomb spin network with two antiferromagnetic exchange interactions $J_1$ and $J_2$ provides a better fit to the susceptibility than the previously proposed spin-chain model. Two sets of the coupling parameters $J_1:J_2=1:0.45$ with $J_1=5.79(1)$~meV and $0.65:1$ with $J_2=6.31(1)$~meV yield equally good fits down to $\sim T_N$. Below $T_{N}$, weak ferromagnetism due to spin canting is observed. The canting is caused by the Dzyaloshinskii-Moriya interaction with an estimated $bc$-plane component $\left|D_p\right|$ $\simeq0.14J_1$. Neutron diffraction reveals that the $S=1/2$ Cu$^{2+}$ spins antiferromagnetically align in the $Fd'd'2$ magnetic space group. The ordered moment of 0.93(9)~$\mu_B$ is predominantly along the crystallographic $a$-axis. 

\end{abstract}
	
	
\maketitle 
\section{Introduction}
Low-dimensional quantum magnetism has attracted much interest from both theoretical and experimental condensed matter physicists for many decades.\cite{Bonner1964,Sachdev2008} It is known that in a one dimensional $(1D)$ antiferromagnetic system, long range order is absent even at zero temperature,\cite{Bethe1931} leading to various fascinating magnetic ground states and phenomena at low temperatures such as the spin-Peierls state in CuGeO$_{3}$~\cite{Hase1993, Riera1995} and TiOCl,\cite{Shaz2005,Abel2007} the singlet ground state in the alternating spin-chain system (VO)$_{2}$P$_{2}$O$_{7}$,\cite{Garrett1997} the fractional spinon excitations in CuSO$_{4}\cdot$5D$_{2}$O, \cite{Mourigal2013} and the Bose-Einstein condensation (BEC) of magnons in the double spin-chain system TlCuCl$_{3}$,\cite{nikuni} in which weakly interacting dimers are formed at low temperatures and BEC is realized as the field-induced $3D$ magnetic ordering. However, some approximate $1D$ antiferromagnets exhibit long-range order with a remnant of quantum fluctuations in a form of quantum renormalization of spin waves.\cite{desCloizeaux,Endoh}  Hence, in order to apprehend diverse physics of these low-dimensional quantum magnets, it is crucial to identify a spin network and relevant underlying interactions that consequently cause magnetic ordering and govern spin dynamics. 

\begin{figure}
	\begin{center}
		\includegraphics[width=8.2cm]{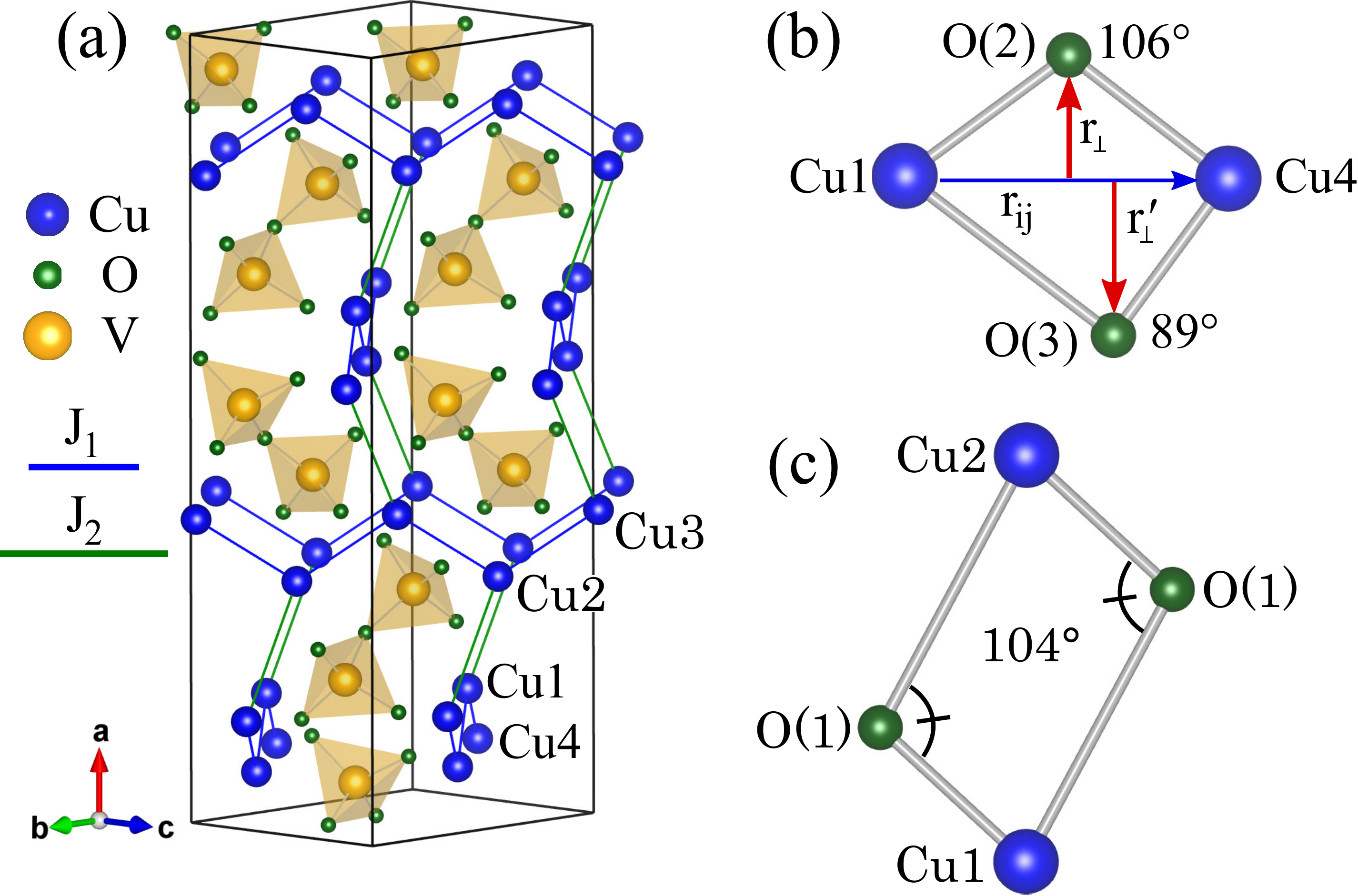}
		\caption{(Color online) (a) Crystal structure of $\alpha$-Cu$_{2}$V$_{2}$O$_{7}$. The chains of edge sharing CuO$_{5}$ polyhedra are along [011] and [01$\bar{1}$] with two corner-sharing V$_2$O$_{7}$ tetrahedra intercalated between the chains. Blue ($J_1$) and green ($J_2$) bonds represent the antiferromagnetic couplings between the nearest-neighbor, intra-chain spins and next-nearest-neighbor, inter-chain spins, respectively. (b) The DM interactions, the directions of which can be determined from the cross product of $\vec{r}_{\perp}\times\vec{r}_{ij}$ and $\vec{r}'_{\perp}\times\vec{r}_{ij}$ for two inequivalent Cu1--O(2)--Cu4 and Cu1--O(3)--Cu4 bonds, respectively, yield a non-zero value. (c) On the other hand, the equivalent Cu1-O(1)-Cu2 pathways connecting the inter-chain spins give rise to a net vanishing DM interaction. The structures are visualized using {\it VESTA}.\cite{Vesta}\label{fig1}}
	\end{center}
\end{figure}

Among various compounds, copper-based oxides with Cu$^{2+}$ ions (3$d^9$), titanium-based oxides with Ti$^{3+}$ ions (3$d^1$), and vanadium-based oxides with V$^{4+}$ ions (3$d^1$) are generally considered as a good realization of the low-dimensional spin-1/2 system. Copper divanadate Cu$_{2}$V$_{2}$O$_{7}$ is a promising realization of the low-dimensional spin-1/2 system. There are two polymorphs of Cu$_{2}$V$_{2}$O$_{7}$ that are generally found in a stable phase, namely the $\alpha$-phase and the $\beta$-phase with a structural phase transition at around $605 ^\circ$C.\cite{Clark1977} Other forms such as the $\gamma$- and $\beta'$-phases were also reported as complex and unstable high-temperature phases.\cite{Clark1977, Krivovichev2005, Krasnenko2008} The $\alpha$-phase of Cu$_{2}$V$_{2}$O$_{7}$ crystallizes in the orthorhombic system of space group {\sl Fdd2} with $a$ = 20.645(2)~\AA, $b$ = 8.383(7)~\AA, and $c$ = 6.442(1)~\AA.\cite{Calvo1975, Robinson1987} Structurally, Cu$^{2+}$~ions that are surrounded by five oxygen ions appear to form a chain of edge sharing CuO$_{5}$ polyhedra while the nonmagnetic V$^{5+}$ ions, each of which is surrounded by four oxygen ions to form V$_2$O$_7$ double corner-sharing tetrahedra, are intercalated between the chains as shown in Fig.~\ref{fig1}(a). Within the $bc$-plane, the $S=1/2$ spins of Cu$^{2+}$ ions form a zigzag chain and interact with their nearest neighbors via two inequivalent Cu--O--Cu pathways. The distance between the two copper ions Cu1 and Cu4 connected by the blue bonds in Fig.~\ref{fig1}(a) is approximately 3.1~\AA.  An intra-chain interactions between these two coppers are formed by the bridging of Cu1(Cu2) with Cu4(Cu3) along [011]([01$\bar{1}$]) via O(2) and O(3). The deviation of the Cu--O(2)--Cu angle from 90$^\circ$ (Fig.~\ref{fig1}(b)) leads to a preferable and strong antiferromagnetic interaction along the chain.\cite{Goodenough,*Kanamori,*Anderson} In the previous study, this system has been proposed as a realization of a zigzag spin-chain model.\cite{Pommer2003} However, the possibly strong next-nearest-neighbor interactions between the Cu$^{2+}$ spins, which link the zigzag chains along the crystallographic $a$-axis as shown by the green bonds in Fig.~\ref{fig1}(a), suggest a highly anisotropic and non-trivial spin network with three coordinate spins for each site that resembles a disconnected honeycomb (helical-honeycomb) lattice when viewed along the $b$-axis (Fig.~\ref{fig9}). The next-nearest-neighbor bond of length 4.0~{\AA}, which is slightly longer than the nearest-neighbor bond, is formed by the bridging of Cu1(Cu3) with Cu2(Cu4) via two equivalent O(1) ions (Fig.~\ref{fig1}(c)) with the Cu--O(1)--Cu angle of 104$^\circ$, possibly leading to an antiferromagnetic interaction that could be comparable to the intra-chain interaction. This sizable inter-chain interaction that enables the forming of the helical-honeycomb lattice is not surprising. Using first-principle calculations on its cousin phase $\beta$-Cu$_{2}$V$_{2}$O$_{7}$, Tsirlin {\it et al.}\cite{Tsirlin} also suggested that the $\beta$-phase could be better described by a honeycomb lattice than by a spin-chain model. The coordination number of three is the lowest for any $2D$ lattices and only a $1D$ chain has a lower coordination number.  Therefore, the helical-honeycomb model could share many properties unique to the low-dimensional magnets even though the lattice extends in three dimensions. 
	
Magnetic susceptibility and heat capacity on a powder sample reveal that the antiferromagnet $\alpha$-Cu$_{2}$V$_{2}$O$_{7}$ magnetically orders at a N\'eel temperature $T_N$ of 34~K.\cite{Ponomarenko2000, Pommer2003} The magnetic ground state of $\alpha$-Cu$_{2}$V$_{2}$O$_{7}$ was previously proposed to be a canted antiferromagnetic spin-chain.\cite{Ponomarenko2000, Touaiher2004, Pommer2003} The canting of spins is due to Dzyaloshinskii-Moriya (DM) interations, which give rise to the presence of weak ferromagnetism below $T_N$. However, the magnetic structure and spin dynamics of $\alpha$-Cu$_{2}$V$_{2}$O$_{7}$ have not been studied. Among the various probing techniques, neutron scattering is the most powerful at revealing the microscopic properties of such magnetic materials; however, a good quality large single crystal is required. Here we report the first detailed study of magnetic properties on single-crystal $\alpha$-Cu$_{2}$V$_{2}$O$_{7}$ using magnetization, Quantum Monte Carlo (QMC) simulations, and neutron diffraction measurements.
	
The paper is organized as follows. In Sec.~\ref{exp}, we briefly discuss single-crystal growth and experimental techniques used to characterize and study magnetic properties of the compound.  The results in Sec.~\ref{results} are divided into three parts, where x-ray diffraction (Sec.~\ref{sec:xray}), magnetization and QMC simulations (Sec.~\ref{sec:susceptibility}), and neutron diffraction (Sec.~\ref{sec:neutron}) will be discussed. Finally, we end with the summary in Sec.~\ref{summary}. 
	
\section{Experimental details}\label{exp}
Prior to the single-crystal growth, powder Cu$_{2}$V$_{2}$O$_{7}$ was prepared from high purity CuO and V$_{2}$O$_{5}$. The chemicals were dehydrated and weighed with stoichiometric ratio and then ground thoroughly with ethanol. The mixture was then calcined at $500^\circ$C for 24 hours. The obtained powder was inserted into a quartz tube. The bottom end of the tube was shaped into a taper for seed selection while the top end was tightly closed with silica wool. The sample was melted in air at $850^\circ$C for 10 hours to ensure homogeneity and then lowered through a constant temperature gradient of about $40^\circ$C/cm inside a vertical Bridgman furnace at a speed of 1 cm/day. The sample was finally cooled from $700^\circ$C to room temperature at a rate of $5^\circ$C/min. Single crystals were extracted from the quartz tube by mechanical separation. A pure-phase powder sample of $\alpha$-Cu$_{2}$V$_{2}$O$_{7}$ was also synthesized by the standard solid state reaction and used in powder neutron diffraction measurements.
	
Small pieces of the crushed crystals were collected and ground for powder x-ray diffraction measurements using Cu$K\alpha$ radiation. The results were fit using the Rietveld method in {\sl FullProf}.\cite{fullprof} To confirm the crystal structure, single-crystal x-ray diffraction data were collected at room temperature using a Bruker X8 APEX CCD diffractometer with Mo$K\alpha$ radiation. The refinements were done using the software {\sl ShelXle}.\cite{shelxle} The magnetization $M$ of the single-crystal sample was measured to the lowest temperature of 2 K using a superconducting quantum interference device (MPMS-XL, Quantum Design). 

QMC simulations with the \textsc{loop} algorithm\cite{loop,*Evertz2003} were performed using the simulation package \textsc{alps}.\cite{alps} The magnetic susceptibility was calculated on a cluster of 100 spins for the spin-chain model and up to 432 spins (27 unit cells) for the helical-honeycomb model with a periodic boundary condition in the temperature range of $0.01 \leq t \leq 5$ ($t=k_{B}T /J$) using 100~000 sweeps for thermalization and 500~000 Monte Carlo steps after thermalization.

To check crystallinity quality and investigate the magnetic transition, single-crystal neutron diffraction was performed at the BT7 Double Focusing Thermal Triple Axis Spectrometer\cite{BT7} at NIST Center for Neutron Reseach, USA, on a single crystal with a mosaic of $0.8^\circ$. Elastic neutron scattering were performed at 2.5~K and 50~K using fixed incident energies of 14.7~meV and 30.5~meV. The position-sensitive detector (PSD) was employed in a two-axis mode with open~\---~80$'$~\---~sample~\---~80$'$(radial)~\---~PSD horizontal collimations to map out the broad reciprocal space in the $(hk0)$ scattering plane (Fig.~\ref{fig6}(a)). A detailed investigation of the nuclear and magnetic Bragg reflections were performed using a triple-axis mode with a single detector and the horizontal collimations of open \---~80$'$~\---~sample~\---~80$'$~\---~120$'$.  For all diffraction measurements, one pyrolytic graphite (PG) filter and two  PG filters were placed along the incident and scattered beams, respectively, to reduce higher-order neutron contamination.  Neutron diffraction on the powder sample were performed at the high-resolution neutron diffractrometer Echidna, ANSTO, Australia using neutrons with wavelength 2.44~\AA (13.7 meV). The data were collected at 3~K and 40~K to extract the magnetic scattering. Additional single-crystal neutron diffraction measurements were carried out at 4~K and 50~K using the Laue diffractometer Koala, ANSTO, Australia. The Laue data images were processed using the {\sl LaueG} software.\cite{piltz} 
 	
\section{Results and Discussion}\label{results}

\subsection{\label{sec:xray}X-ray diffraction}

\begin{figure}
\begin{center}
\includegraphics[width=8cm]{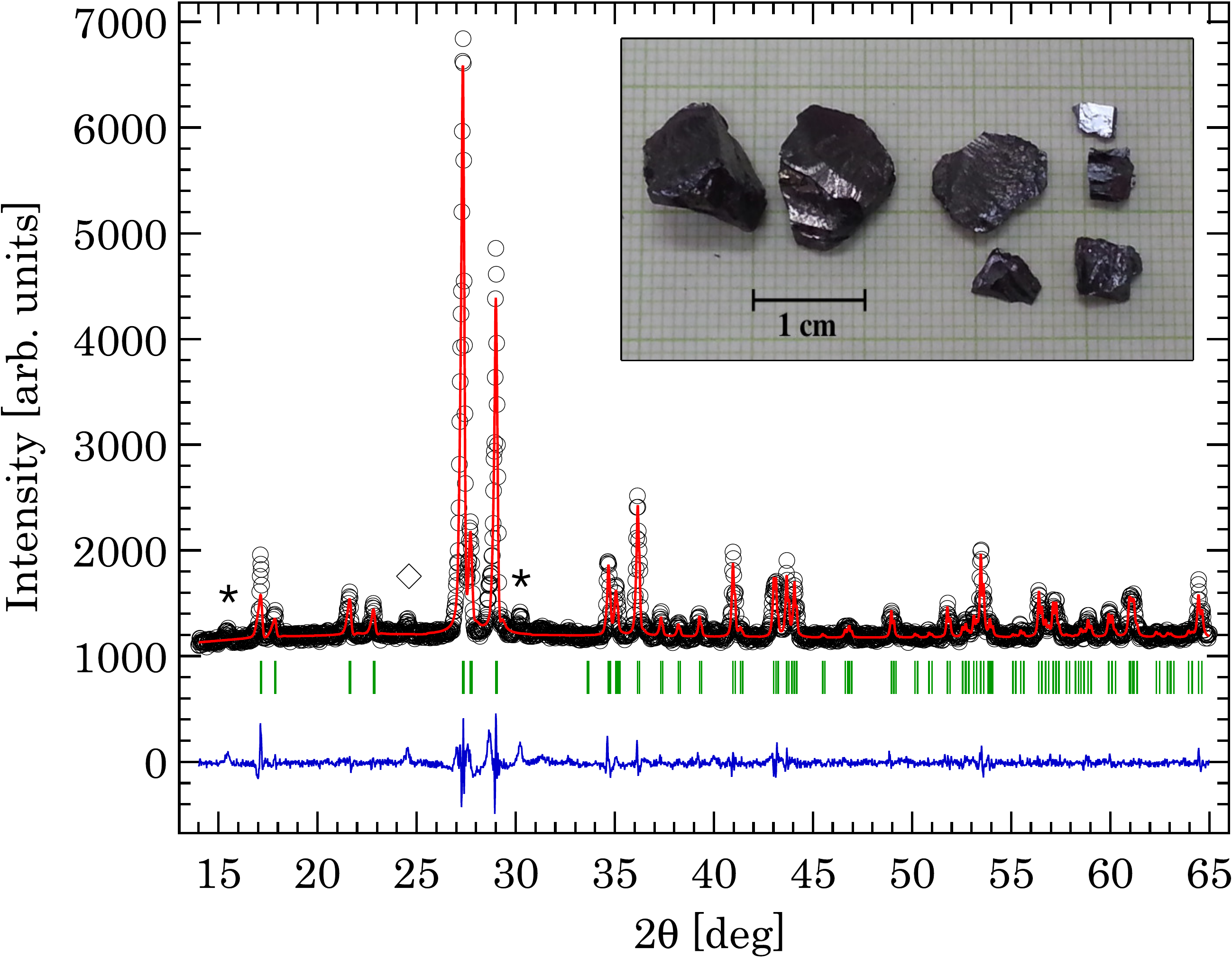}
\caption{(Color online) Powder x-ray diffraction was measured on crushed single crystals, black for the observed data, red for the calculated pattern, blue for the difference, and green for the Bragg positions. The symbols $\star$ and $\diamond$ indicate Cu$_{0.64}$V$_{2}$O$_{5}$ and $\beta$-Cu$_{2}$V$_{2}$O$_{7}$ impurities, respectively. The inset shows a photograph of the $\alpha$-Cu$_{2}$V$_{2}$O$_{7}$ single crystals. \label{fig2}}
\end{center}
\end{figure}

The obtained single crystals are shown in the inset of Fig.~\ref{fig2}. The largest crystal was about 1~$\times$~1~$\times$~0.5~cm$^3$ with a mass of 1.4 g. The naturally cleaved facet is the (1,0,0) plane. Small crystals were collected for the x-ray diffraction measurements while large crystals were allocated for the neutron scattering experiments. The result of the powder x-ray diffraction (Fig.~\ref{fig2}) shows that the major phase of the crystals is $\alpha$-Cu$_{2}$V$_{2}$O$_{7}$ ($\sim$95\%) with a small amount of impurities that can be identified as $\beta$-Cu$_{2}$V$_{2}$O$_{7}$ and Cu$_{0.64}$V$_{2}$O$_{5}$.\cite{Christian1974} We note that no trace of these impurities is detected in the powder sample (Fig.~\ref{fig7}). The lattice constants obtained from the refinement are $a$~=~20.678(6)~\AA, $b$ = 8.405(2)~\AA, and $c$ = 6.446(2)~\AA, which are in good agreement with those reported in Ref.~\onlinecite{Robinson1987}. The impurities are still present after several crystal growth attempts with different cooling conditions, which are an important factor to control the ratio of $\alpha$ and $\beta$ phases.\cite{Slobodin2009, Clark1977} We found that the fraction of $\beta-$phase increases with an increasing cooling rate. Hence it is crucial to slowly cool the sample through the phase transition temperature at 605$^\circ$C to avoid a mixture of the two polymorphs. It should be noted that these impurities comprise only small percentage, are most likely in a powder form, and hence will not mislead the interpretation of the neutron scattering data (Figs.~\ref{fig6} and \ref{fig8}). Room-temperature single-crystal x-ray diffraction was performed on the crystal with a few hundreds of micrometers in size. The data were refined against space group $Fdd2$ with the previously reported lattice parameters\cite{Calvo1975} yielding the agreement factor {\sl R$_{1}$}~=~0.039 for 1031 reflections with $F_{obs} < 4\sigma(F_{obs}$). The result is shown in Table~\ref{param}. The refinement result from single-crystal neutron diffraction measured at 50~K, which will be discussed later, is also shown in the table for comparison.

\subsection{\label{sec:susceptibility}Magnetic susceptibility}

\begin{figure}
\includegraphics[width=8.5cm]{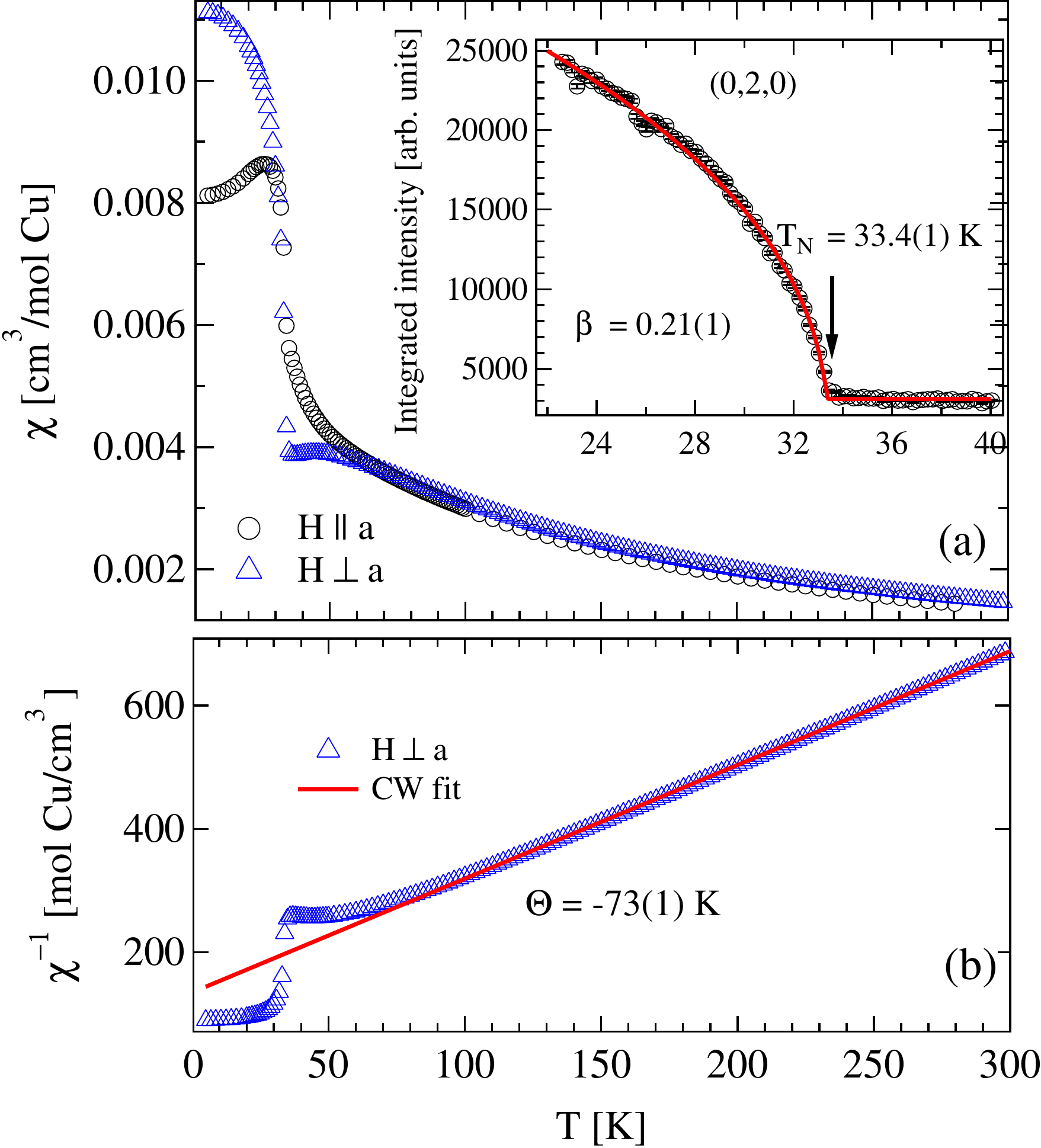}
\caption{(Color online) (a) Magnetic susceptibility was measured with the applied field of 1 T parallel ($H {\parallel} a$) and perpendicular ($H {\perp} a$) to the crystallographic $a$-axis. The inset shows an order parameter as a function of temperature measured by neutron scattering at the magnetic Bragg reflection (0,2,0). (b) The Curie-Weiss law (red) is fit to the susceptibility with $H {\perp} a$.\label{fig3}}
\end{figure}

To investigate the magnetic transition on single-crystal $\alpha$-Cu$_{2}$V$_{2}$O$_{7}$, we measured the magnetic susceptibility ($\chi = M/H$) as a function of temperature when the applied magnetic field of 1 T was parallel and perpendicular to the $a$-axis in the zero-field-cooled mode.  In Fig.~\ref{fig3}(a) the susceptibility exhibits clear anisotropy at low temperatures and shows a sharp transition at $\simeq$~33~K in agreement with $T_{N}$ = 33.4(1)~K obtained from the order parameter measured by neutron diffraction on a single crystal at BT7 (the inset). The N\'eel temperature is consistent with that obtained from the susceptibility measured on our powder sample (not shown), as well as with those from the previous powder-sample studies.\cite{Ponomarenko2000, Pommer2003}

Above 100~K, a linear fit of $\chi^{-1}$ as a function of $T$ to the Curie-Weiss law $\left(\displaystyle\chi = \frac{C}{T - \Theta}\right)$ as shown in Fig.~\ref{fig3}(b) gives a Curie-Weiss constant $C$ = 0.545(2) cm$^3$K/mol~Cu and a Curie-Weiss temperature $\Theta = -73(1)$~K. The negative Curie-Weiss temperature suggests that the dominant exchange interactions are antiferromagnetic.  From the Curie-Weiss constant, the calculated effective moment $\mu_{eff}=\sqrt{3k_{B}C/N_{A}}=2.087(4)~\mu_{B}$ is obtained. This value is slightly higher than the spin-only value of $\mu_{eff}=g\mu_{B}\sqrt{S(S+1)}=1.73~\mu_{B}$ for $g=2$ and $S=1/2$ for Cu$^{2+}$ ions. The order of frustration defined by $f=\left|\Theta/T_{N}\right|$ is $\simeq2.2$, which suggests that the spin interactions are not strongly frustrated (a typical value for strongly frustrated systems is $f>10$). Assuming the mean field approximation, one can calculate the antiferromagnetic exchange interaction from $\Theta=-zJ_{CW}S(S+1)/3k_B$, where $z$ is a number of the coordinate spins and $S=1/2$.  The calculations give $J_{CW}=12.6(2)$~meV for the spin-chain model with $z=2$ and $8.4(2)$~meV for the helical-honeycomb lattice, in which each spin has three coordinate spins, two along the zigzag chain and one between the chains (Figs.~\ref{fig1}(a) and~\ref{fig9}(a)), giving $z=3$.
 
Below $T_N$, when the magnetic field is applied parallel to the $a$-axis, a small cusp can be observed at the magnetic ordering transition. This cusp is a signature of an antiferromagnetic transition and suggests that the spins align anti-parallel along the crystallographic $a$-axis. On the other hand, when the field is applied perpendicular to the $a$-axis, the susceptibility shows a broad maximum around 50~K suggesting a rise of short-range correlations, typical for low-dimensional magnets, before an abrupt increase indicative of long-range ordering at lower temperatures. The weak ferromagnetism for $H~{\perp}~a$ below $T_{N}$ is a result of small spin canting due to the DM interactions. Hence, to first approximation, the spin Hamiltonian can be described by 
\begin{equation}
{\cal H}=\sum_{\langle i,j\rangle}\left[J_{ij}\vec{S}_i\cdot\vec{S}_j+\vec{D}_{ij} \cdot(\vec{S}_{i} \times\vec{S}_{j})\right], 
\end{equation}
where $J_{ij}$ denotes the exchange couplings representing the nearest-neighbor, intra-chain $J_{1}$ and next-nearest-neighbor, inter-chain $J_2$ interactions as shown by the blue and green bonds in Fig.~\ref{fig1}(a), respectively. {\bf D$_{ij}$} is the DM vector whose strength is proportional to spin-orbit coupling and scaled with the exchange interaction between $\vec{S}_{i}$ and $\vec{S}_{j}$. The DM interaction is present in $\alpha$-Cu$_{2}$V$_{2}$O$_{7}$ for the Cu1--Cu4 bond (Fig.~\ref{fig1}(b)) since there is no inversion center between the magnetic Cu$^{2+}$ ions.\cite{moriya}  The magnitude of spin canting $\eta$ resulted from the DM interactions was estimated to be $\simeq 2^\circ$ from the previous magnetization measurements on the powder sample.\cite{Pommer2003}

\begin{figure}
	\includegraphics[width=8.5cm]{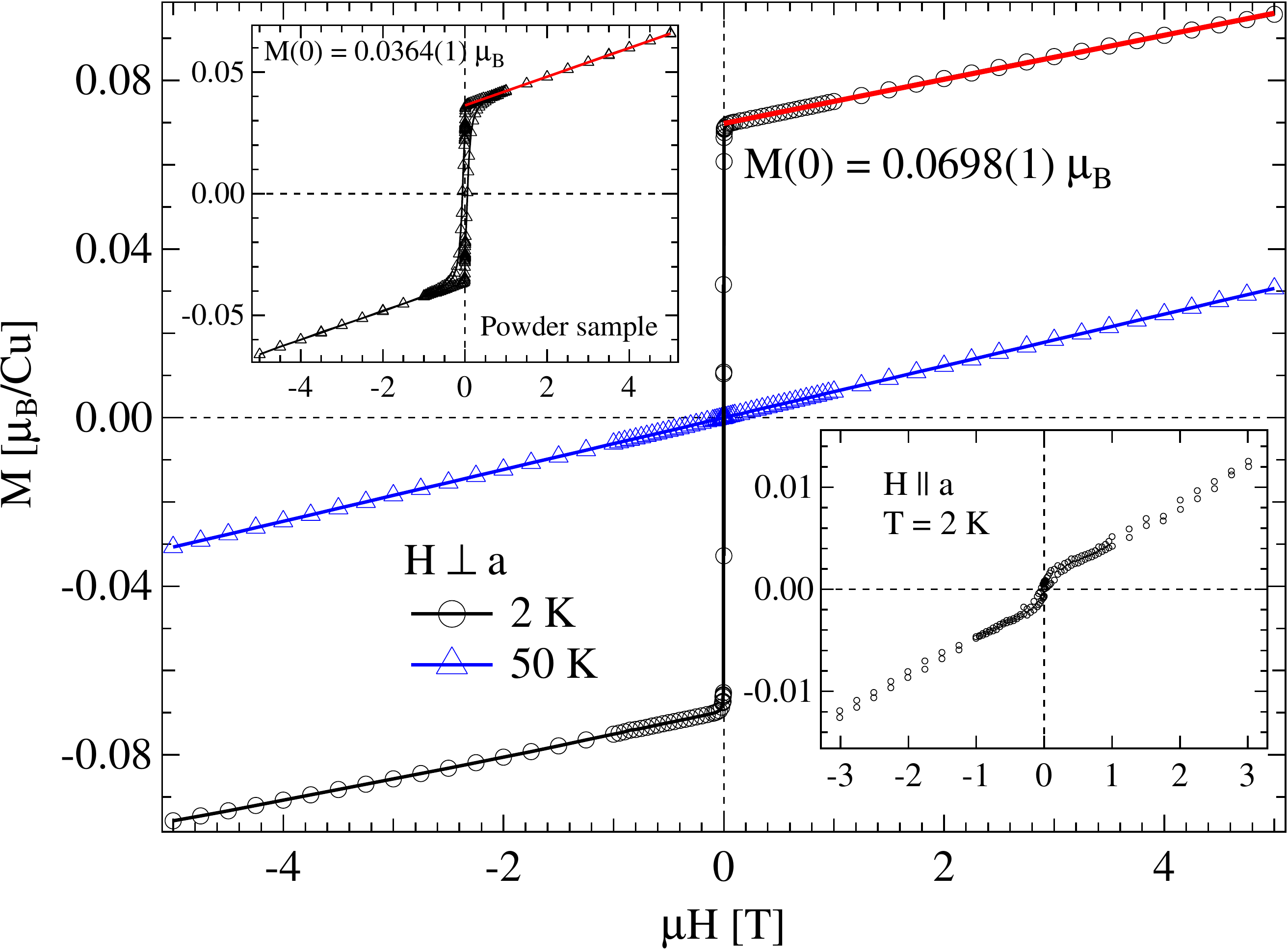}
	\caption{(Color online) Magnetization as a function of magnetic field was measured on single-crystal $\alpha$-Cu$_{2}$V$_{2}$O$_{7}$ with $H \perp a$ at T = 2 K and 50 K. The lower inset shows the data for $H \parallel a$. The upper inset shows the magnetization as a function of magnetic field measured on a powder sample at T = 5~K.\label{fig4}}
\end{figure}

To determine the spin canting and DM parameter on the single-crystal sample, the magnetization measurements as a function of magnetic field up to a maximal field of 5~T were performed with two orthogonal magnetic field orientations, {\it i.e.} $H~{\parallel}~a$ and $H~{\perp}~a$. Figure~\ref{fig4} shows the magnetization measurement on the single crystal with the applied magnetic field perpendicular to the $a$-axis. Above $T_{N}$ ($T=50$~K) the magnetization is linear throughout the measuring field range. However, below $T_{N}$ ($T=2$~K) starting from zero applied field the magnetization sharply rises to a finite value with only a slight increase in magnetic field before attaining the same linear response as that observed in the 50~K data. The rapid increase of the magnetization, which is indicative of weak ferromagnetism due to the spin canting, is not observed for $H~{\parallel}~a$ as shown in the lower inset. The small kink around zero field is most likely due to slight misalignment. The same measurements on the power sample show a similar weak ferromagnetic component as shown in the upper inset. However, the jump is less sharp and about a factor of two smaller due to powder average. For the single-crystal data, a hysteresis loop, which is typical for ferromagnetism, is not clear with a very small coercive field ($< 10$ Oe, which is the resolution of the measurements), but it is more pronounced for the powder data, possibly due to the powder average over all orientations, which could broaden the magnetization reversal. 

Quantitatively, the red line in Fig.~\ref{fig4} (and in the upper inset for the powder sample) denotes a linear fit to the magnetization for $H \geq$ 5000 Oe and is extrapolated to intercept the $y$-axis to obtain the value of $M(0)$, the canted moment at zero field.  From the single-crystal (powder) data, $M(0)$~=~0.0698(1)~$\mu_{B}$ (0.0364(1)~$\mu_{B}$) is obtained. On the other hand, with $H~{\parallel}~a$ the value of $M(0)$ is approaching zero as expected since the predominant spin component is anti-parallel along the $a$-axis and the canted moments only stay within the $bc$-plane.  We note that there  exist two sets of the zigzag chains along $[011]$ and $[01\bar{1}]$ on the alternating planes.  These chains are about 75$^\circ$ with respect to each other (Fig.~\ref{fig9}(b)).  We will assume that for the single-crystal data the measured canted moment $M(0)$ is a saturated value, where the canted moments on both sets of the chains are aligned along the direction of the applied magnetic field. The canting angle can be calculated from the relation $\displaystyle\eta =$ sin$^{-1}\left(\frac{M(0)}{g\mu_{B}S}\right)$. Given the expected spin-only $S = 1/2$, $g = 2$, and $M(0)$ from the single-crystal data, the canted moment of $\eta = 4.0^\circ$ is obtained.  

The direction of the DM vector $\vec{D}_{ij}$ is determined by $\vec{r}_{\perp}\times\vec{r}_{ij}$, where $\vec{r}_{ij}$ is a unit vector connecting the spins $\vec{S}_{i}$ and $\vec{S}_{j}$, and $\vec{r}_{\perp}$ is perpendicular to $\vec{r}_{ij}$ and points toward the oxygen ligand as shown in Fig.~\ref{fig1}(b). The two pathways Cu1--O(2)--Cu4 and Cu1--O(3)--Cu4 are structurally inequivalent giving rise to non-compensating DM interactions. We note that the DM interactions arising from the two equivalent Cu1--O(1)--Cu2 bonds (Fig.~\ref{fig1}(c)) compensate each other resulting in the vanishing DM vector. Although the resulting $\vec{r}_{\perp}\times\vec{r}_{ij}$ does not restrict the DM vector to a specified high-symmetry plane, the cross product of a spin pair, $\vec{S}_{i}\times\vec{S}_{j}$, constrains the relevant component of the DM vector, which causes the canting, to be only within the $bc$-plane. We will later show in Sec.~\ref{sec:neutron} that the canted moments in the $bc$-plane are parallel, and hence the component of $\vec{S}_{i}\times\vec{S}_{j}$ along the $a$-axis vanishes. If only the relevant interaction $J_{1}$ between Cu1 and Cu4 is taken into account, the in-plane DM parameter $D_p$ can be related to the canting angle $\eta$ and $J_{\text{1}}$ through the following relation: 
\begin{eqnarray}
\tan(2\eta) =\left|\frac{D_p}{J_{\text{1}}}\right|,\label{eq2}
\end{eqnarray}
where the absolute value denotes the undetermined direction of the in-plane DM vector.

\begin{table*}
	\caption{\label{table1}Parameters obtained from the fit of magnetic susceptibility with $H \perp a$ using different lattice models.}
	\centering
	\begin{tabular}{c c c c c}
		\hline \hline
		& Bethe ansatz& \multicolumn{3}{c}{ QMC (this work)} \\
		& calculation~\cite{Johnston}  & uniform & \multicolumn{2}{c}{helical-honeycomb}\\
		&~~~~~(coupled spin-chain)~~~~~&~~~~~spin-chain~~~~~&~~~~~$J_{1} > J_{2}~~~~~$ &~~~~~$J_{1} < J_{2}~~~~~$\\
		\hline
		$J_{1}$ [meV] & 5.95(2) & 5.95(2) & 5.79(1) &  4.10(1)\\
		$J_{2}$ [meV] & $\simeq$1.42 & $-$ & 2.61(1) & 6.31(1)\\
		$g$ & 2.16(1) & 2.16(1) & 2.24(1) & 2.25(1)\\
		~~~~~$\chi_{\z}~\left[10^{-4}\text{ cm}^{3}/\text{mol Cu}\right]$~~~~~ & 1.68(7) & 1.69(7) & 0.94(6) & 0.89(3)\\
		$\left|D_{p}\right|$ [meV] & 0.836(3) & 0.836(3) & 0.814(1) & 0.576(1)\\[0.1cm]
		\hline \hline
	\end{tabular}
\end{table*}

\begin{figure}
	\includegraphics[width=8cm]{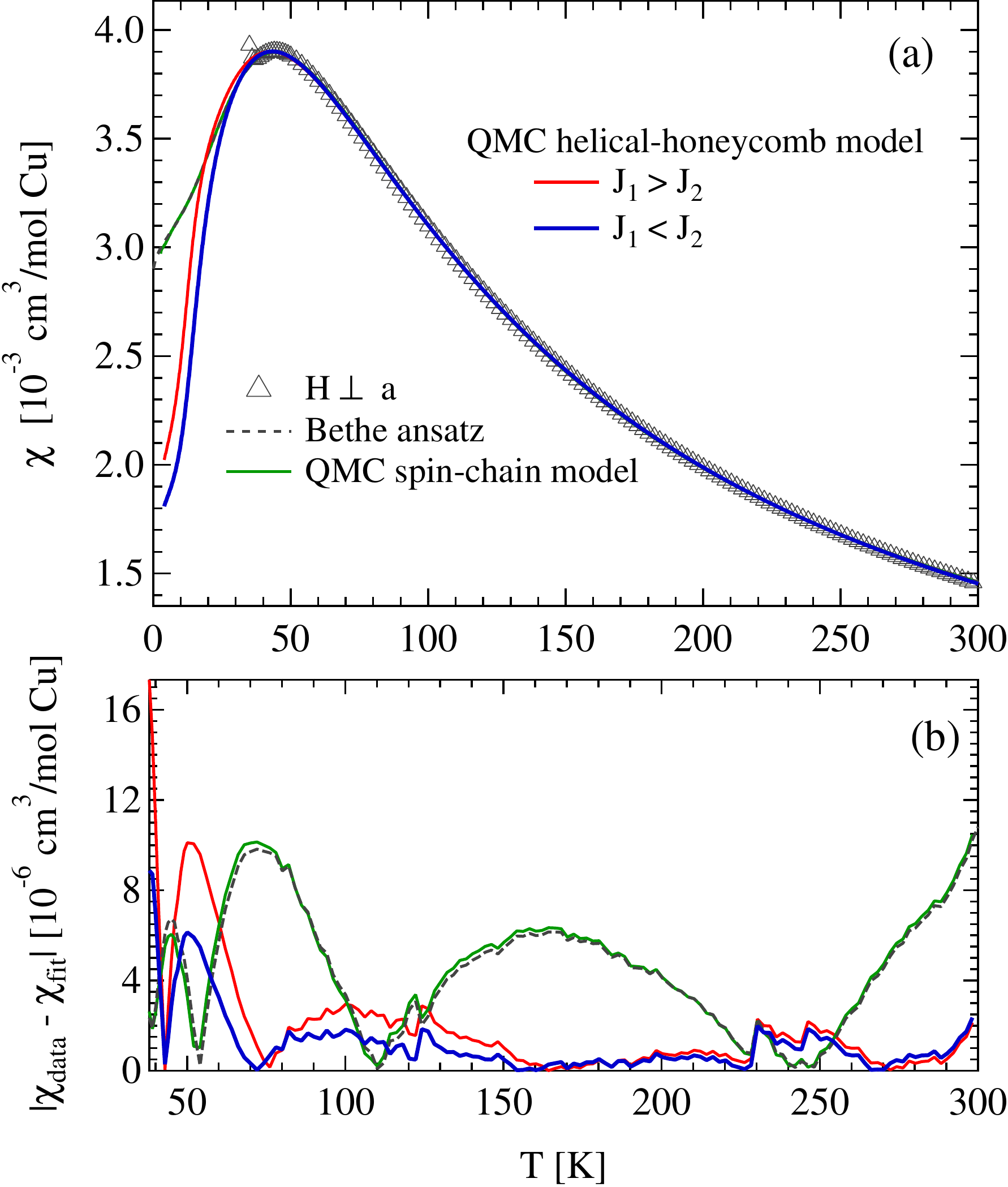}
	\caption{(Color online) (a) Fit of magnetic susceptibility with $H\perp a$ to the Bethe ansatz numerical result (black-dashed line), QMC simulations on the uniform spin-chain model (green line), QMC simulations on the helical-honeycomb model with $J_{1}>J_{2}$ (red), and with $J_{1}<J_{2}$ (blue). (b) Residual curves ($\left| \chi_{\text{data}} - \chi_{\text{fit}} \right|$) between the data and fit curves of different lattice models.\label{fig5}}
\end{figure}

\begin{figure}
\includegraphics[width=8.5cm]{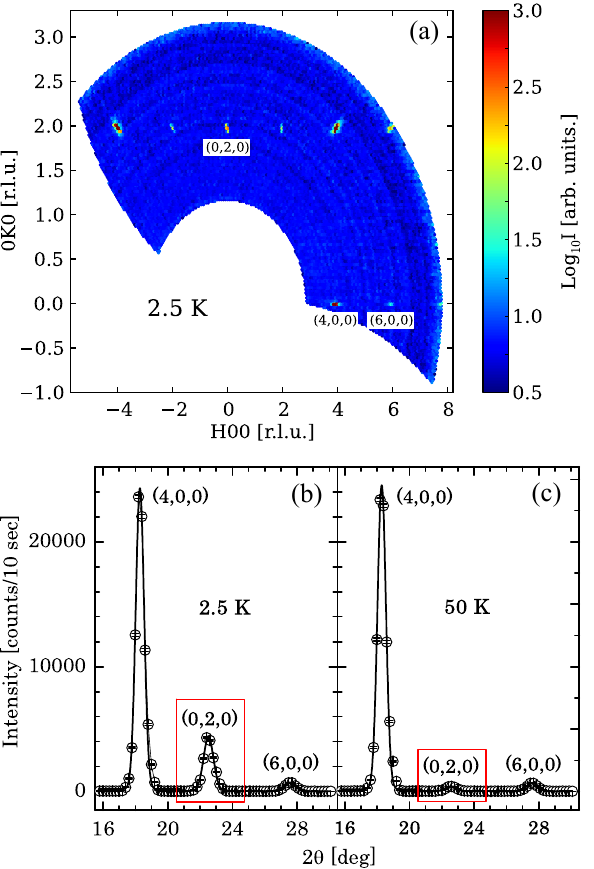}
\caption{(Color online) (a) An intensity map of single-crystal $\alpha$-Cu$_{2}$V$_{2}$O$_{7}$ was measured in the $(h,k,0)$ plane at 2.5 K. The intensity is presented in a log scale. (b) and (c) show the $\theta-2\theta$-scans around the (4,0,0), (0,2,0), and (6,0,0) Bragg reflections at 2.5~K and 50~K, respectively. The solid line is a guide to the eye. Error bars represent one standard deviation throughout the article.
\label{fig6}}
\end{figure}

In order to estimate the exchange couplings $J_{1}$ and $J_{2}$, we first analyze magnetic susceptibility based on a weakly coupled spin-chain model ($J_{2} \ll J_{1}$). We reconsider the magnetic susceptibility as a function of temperature shown in Fig.~\ref{fig3} and fit the data to the result of the high-accuracy numerical Bethe ansatz calculations\cite{Johnston} $\chi_{\text{\tiny \b}}(T)$ for the $1D$ Heisenberg antiferromagnetic system.  A fit of the measured magnetic susceptibility (Fig.~\ref{fig5}) to
\begin{eqnarray}
\chi(T) = \chi_{\z} + \chi_{\text{\tiny \b}}(T),\label{chi_eq}
\end{eqnarray}
where $\chi_{0}$ is a temperature-independent susceptibility background, yields $\chi_{0}$ = 1.68(7) $\times$ 10$^{-4}$ cm$^3$/mol Cu, the intra-chain coupling $J_{1}=5.95(2)$~meV, and the Land\'e $g$-factor $g$ = 2.16(1).  The Curie-Weiss exchange coupling $J_{CW} = 12.6$ meV obtained from the mean field approximation for a number of coordinate spins $z=2$ is substantially larger than $J_{1}$ obtained from the susceptibility fit to the spin-chain model ($\left|J_1-J_{CW}\right|/J_{CW}\simeq0.53$) suggesting that the coordination number must be greater than two. Therefore, the inter-chain interaction is non-negligible. Using Schulz's calculation of the inter-chain interaction for parallel chains in th1e mean field approximation, $\left|J_{\text{inter}}\right|=\left|J_{2}\right|=T_{N}/\left(1.28\sqrt{\text{ln}(5.8J_{\text{intra}}/T_{N})}\right)$,\cite{Schulz1996} where $J_{\text{intra}}$ is an intra-chain interaction, and applying it to $\alpha$-Cu$_2$V$_2$O$_7$ with $T_{N}$~=~33.4~K and $J_{\text{intra}}=J_{1}=5.95$~meV, we obtain $\left|J_{2}\right| \simeq$ 1.42~meV. Even though the sign of $J_{2}$ cannot be determined from this calculation, the discrepancy between $J_{CW}$ for $z = 2$ and $J_{1}$ suggests that the inter-chain interaction is dominantly antiferromagnetic and hence $J_{2}>0$. Neutron diffraction, which will be discussed in the next section, reveals antiferromagnetic alignment of spins in the $[011]$ and $[01\bar{1}]$ chains on different $bc$-planes as shown in Fig.~\ref{fig9}. Therefore, the leading antiferromagnetic inter-chain interactions (denoted by green bonds in Fig.~\ref{fig9}(a)), which induce the anti-parallel arrangement of spins along the $a$-axis, link the chains on the adjacent planes giving rise to the helical-honeycomb spin network. The Cu--O--Cu bond angle of this inter-chain interaction, which is greater than 90$^\circ$ (Fig.~\ref{fig1}(c)), is also consistent with the antiferromangetic coupling. The coupling between the parallel chains in the same $bc$-plane is expected to be much smaller because the bonding is not of the Cu--O--Cu type but must be via the V$_2$O$_7$ double tetrahedra. The value of $\left|J_{2}/J_{1}\right|\simeq0.23$ is substantially larger than that measured in other $1D$ systems such as KCuF$_{3}$, Sr$_{2}$CuO$_{3}$, and BaCu$_{2}$Si$_{2}$O$_{7}$, where the value of $\left|J_{\text{inter}}/J_{\text{intra}}\right|$ is $0.001-0.01$,\cite{Kojima1997, Tennant1995, Tsukada1999} suggesting that the magnetism in $\alpha$-Cu$_{2}$V$_{2}$O$_{7}$ is far from being an ideal realization of the $1D$ spin system, and probably invalidating the above analysis as well as other previous studies, which are based on the coupled spin-chain model.\cite{Ponomarenko2000, Touaiher2004, Pommer2003} Furthermore, we note that since the dominating inter-chain interaction is between the $[011]$ and $[01\bar{1}]$ chains, of which the spin coordination number is different from that of the parallel chains, Schulz's results used above could potentially provide an erroneous magnitude of $J_2$.

To obtain the better estimates of $J_1$ and $J_2$, we perform QMC simulations with the \textsc{loop} algorithm on the helical-honeycomb spin network, which shows a disconnected hexagon when viewed along the $b$-axis (Fig.~\ref{fig9}(a) and (c)). The magnetic susceptibility was computed and fit to the data.  In addition, the susceptibility for the uniform spin-chain model was calculated and compared with $\chi_{\text{\tiny\b}}(T)$ (Eq.~\ref{chi_eq}).  The QMC numerical results were first fit using the following formula 
\begin{equation}
\chi^\ast(t) = \frac{1}{4t}{\cal P}^{(q)}_{(r)}(t),\label{pade}
\end{equation}
where $t = k_BT/J_{\text{max}}$ and the Pad\'e approximant is given by
\begin{equation}
{\cal P}^{(q)}_{(r)}(t)=\frac{1+\sum\limits_{n=1}^q N_{n}/t^{n}}{1+\sum\limits_{n=1}^r D_{n}/t^{n}}.
\end{equation}
The coefficient $1/4$ is from $S(S+1)/3$ for $S=1/2$ and the factor $1/t$ ensures that $\chi^\ast(t)$ approaches the Curie law at high temperature. $J_{\text{max}}$ is the maximum of $J_1$ and $J_2$ and the smaller interaction is equal to $\alpha J_{\text{max}}$, where the parameter $\alpha$ is fixed for each QMC simulation.  The numerical parameters $N_{n}$ and $D_{n}$ were obtained up to $q=5$ and $r=6$ from the fit. The resulting Pad\'e approximant for each QMC calculation was then used to fit the measured susceptibility data using
\begin{equation}
\chi(T)=\chi_{\z}+\chi_{\text{\tiny{\q}}}(T),
\end{equation}
where
\begin{equation}
\chi_{\text{\tiny{\q}}}(T)=\frac{N_A\mu_B^2g^2}{k_BJ_{\text{max}}}\chi^\ast\left(k_BT/J_{\text{max}}\right),
\end{equation}
to obtain $J_{\text{max}}$ and $g$. $N_A$, $\mu_B$, and $k_B$ are the Avogadro constant, the Bohr magneton, and the Boltzmann constant, respectively. For the helical-honeycomb model, $\alpha$ is manually adjusted until the best fit is acquired. Figure~\ref{fig5} shows the best fits for both uniform spin-chain and helical-honeycomb models where the obtained fit parameters are summarized in Table~\ref{table1} for comparison. The QMC simulation on the spin-chain model exactly match the result obtained from the Bethe ansatz calculations with the same magnitude of $J_1$, proving the validity to the QMC simulations. For the helical-honeycomb model, two sets of the exchange parameters, for $J_{1} > J_{2}$ $J_{\text{max}}=5.79(1)$~meV, $\alpha = 0.45$, and $g=2.24(1)$ and for $J_{1} < J_{2}$ $J_{\text{max}}=6.31(1)$~meV, $\alpha = 0.65$, and $g=2.25(1)$, fit the data equally well and provide a noticeably better fit than the spin-chain model for most of the temperature range from 300~K down to $\sim T_N$ as shown by the residual analysis in Fig.~\ref{fig5}(b). The obtained average exchange interaction $\bar{J}=\left(2J_1+J_2\right)/3$ for the coupled spin-chain model is much lower than the mean-field value $J_{CW}$ obtained from the Curie-Weiss fit with $\left|\bar{J}-J_{CW}\right|/J_{CW} \simeq 0.47$. Our proposed helical-honeycomb model shows a slightly smaller deviation with $\left|\bar{J}-J_{CW}\right|/J_{CW} \simeq 0.44$ and $\simeq 0.42$ for $J_{1} > J_{2}$ and $J_{1} < J_{2}$, respectively. Additional exchange and anisotropic interactions could affect the value of $\bar{J}$, potentially leading to closer agreement between $\bar{J}$ and $J_{CW}$. Unfortunately, based on combined susceptibility analysis and QMC simulations, we were unable to uniquely identify the leading coupling constant. It was previously proposed that the leading coupling connects Cu1 and Cu4 via $J_1$ forming the zigzag chains along [011] and [01$\bar{1}$] directions (Fig.~\ref{fig1}(a)), but the competition between ferromagnetic coupling via Cu1--O(3)--Cu4 with the bond angle close to 90$^\circ$ and antiferromagnetic coupling via Cu1--O(2)--Cu4 with the bond angle of 106$^\circ$ (Fig.~\ref{fig1}(b)) could render a weaker net antiferromagnetic interaction for $J_1$ than the non-competing antiferromagnetic coupling $J_2$ between Cu1 and Cu2 via two equivalent Cu1--O(1)--Cu2 bridges (Fig.~\ref{fig1}(c)). Further theoretical analyses based on first-principle calculations, which will provide complementary support to our analysis, are desirable in order to identify the leading interaction.

Given $J_{1}$ obtained from the two helical-honeycomb models discussed above and the canting angle $\eta$~=~4.0$^\circ$ from the magnetization data, we estimate the in-plane DM parameter $\left|D_{p}\right|$ using Eq.~(\ref{eq2}) to be $0.814(1)$~meV for the helical-honeycomb model with $J_1 > J_2$ and $0.576(1)$~meV for $J_1 < J_2$, or $\simeq0.14J_1$. The values of $\left|D_{p}\right|$ for all considered lattice models are summarized in Table~\ref{table1}. The magnitude of the DM vector shows good agreement with Moriya's calculation\cite{moriya} as $\Delta g/g\sim 0.12\sim \left|D_p/J_1\right|$, where $g$ is the free electron Land\'e $g$-factor and $\Delta g$ denotes its shift caused by the crystalline environment.  The helical-honeycomb lattice formed by $J_1$ and $J_2$ of comparable strength and the presence of the DM interactions in $\alpha$-Cu$_{2}$V$_{2}$O$_{7}$ induce the magnetically ordered state below the N\'eel temperature. Hence, the long-range order observed in this system does not defy the Mermin-Wagner theorem.\cite{mermin} 

\begin{table}
\caption{\label{param} Refined values of fractional coordinates of $\alpha$-Cu$_{2}$V$_{2}$O$_{7}$ from single-crystal x-ray diffraction measured at room temperature and from single-crystal neutron diffraction measured at 50~K.}
\centering
\begin{tabular}{c c c c c}
\hline \hline
Atom & $x/a$ & $y/b$ & $z/c$ & $U$\\
\hline
\multicolumn{5}{c}{x-ray diffraction}\\
Cu & 0.16572(5) & 0.3646(1) & 0.7545(1) & 0.0143(3)\\
V & 0.19898(5) & 0.4055(1) & 0.2370(2) & 0.0067(3)\\
O(1) & 0.2455(3) & 0.5631(9) & 0.274(1) & 0.022(1)\\
O(2)& 0.1445(3) & 0.4375(6) & 0.0308(8) & 0.0100(9)\\
O(3) & 0.1617(3) & 0.3475(7) & 0.4575(9) & 0.014(1)\\
O(4) & $\frac{1}{4}$ & $\frac{1}{4}$ & 0.156(2) & 0.030(2)\\
\multicolumn{5}{c}{{\sl R$_{1}$} = 0.039, {\sl wR$_{2}$} = 0.095, {\sl GOF} = 1.063}\\
\hline
\multicolumn{5}{c}{neutron diffraction}\\
Cu & 0.16551(2) & 0.36460(5) & 0.7520(1) & 0.0036(1)\\
V & 0.1990(3) & 0.4046(8) & 0.237(2) & 0.0006(9)\\
O(1) & 0.24605(3) & 0.56165(8) & 0.2723(2) & 0.0072(2)\\
O(2) & 0.14428(3) & 0.43776(7) & 0.0286(2) & 0.0040(2)\\
O(3) & 0.16200(3) & 0.34608(9) & 0.4551(2) & 0.0052(2)\\
O(4) & $\frac{1}{4}$ & $\frac{1}{4}$ & 0.1507(3) & 0.0087(3)\\
\multicolumn{5}{c}{{\sl R$_{1}$} = 0.081, {\sl wR$_{2}$} = 0.073, {\sl GOF} = 1.215}\\
\hline \hline
\end{tabular}
\end{table}

\subsection{\label{sec:neutron}Neutron diffraction}

In order to investigate the magnetic structure of the ordered state, we performed elastic neutron scattering to search for magnetic Bragg reflections. Figures~\ref{fig6}(b) and (c) show clear extra scattering intensity below $T_N$ at (0,2,0), indicative of magnetic scattering; small peaks at (0,2,0) and (6,0,0), which are structurally forbidden by symmetry, at 50~K are due to higher-order neutron contamination. On the other hand, the intensities at (4,0,0) and (6,0,0), which are also allowed magnetic Bragg reflections, do not show a significant change below $T_N$.  In neutron scattering, only the spin component that is perpendicular to the momentum transfer contributes to the magnetic scattering intensity, due to the dipole-dipole interaction as described by the geometric factor $\displaystyle\sum_{\alpha,\beta}(\delta_{\alpha,\beta}-\hat{Q}_{\alpha}\hat{Q}_{\beta})$,\cite{Shirane} where $\alpha$ denotes the spin components and $\hat{Q}_{\alpha}$ is the unit vector of ${\mathbf Q}$ along the component $\alpha$.  For $\alpha$-Cu$_{2}$V$_{2}$O$_{7}$ most of the spin component is parallel to the $a$-axis, which is also evidenced by the magnetic susceptibility. Hence the magnetic Bragg reflections $(h,0,0)$ become negligibly small.
 
The integrated intensity of the $(0,2,0)$ magnetic Bragg reflection as a function of temperature (Fig.~\ref{fig3} inset) shows an upturn around 33 K indicating a transition to the antiferromagnetic ordered state, coincident with the jump in the magnetic susceptibility (Fig.~\ref{fig3}(a)). A fit of the order parameter to the power law, $I(T) \propto (1 - T/T_{N})^{2\beta}$, for $24~\textrm{K} < T < 34~\textrm{K}$ gives a critical exponent $\beta$ = 0.21(1) and $T_{N}$ = 33.4(1)~K. The fit value of $\beta$ is typical for low-dimensional magnetic systems.\cite{Tennant1995, Banks2004, Kojima1997}
  
\begin{figure}
\includegraphics[width=8.4cm]{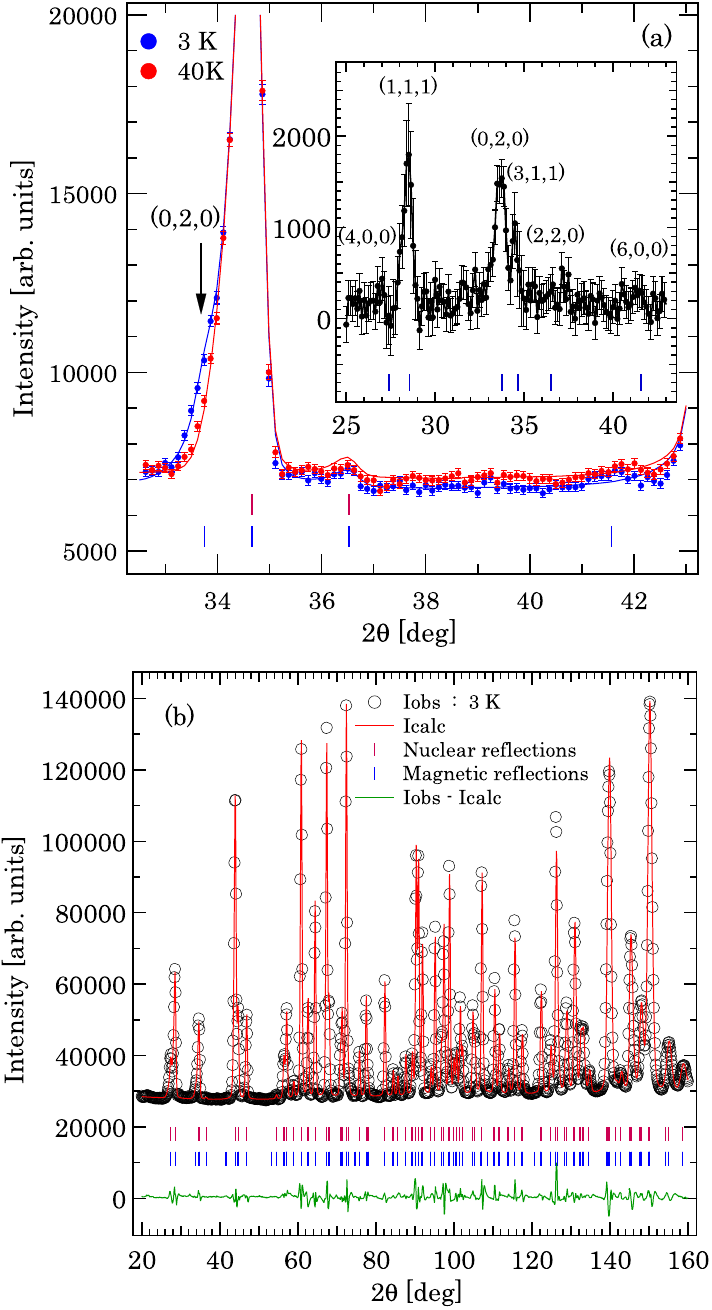}
\caption{(Color online) (a) Neutron diffraction data of powder $\alpha$-Cu$_{2}$V$_{2}$O$_{7}$ at 3 K show the magnetic Bragg scattering at $(0,2,0)$ (indicated by an arrow). The inset shows the intensity difference with allowed magnetic Bragg reflections indexed. (b) The powder neutron diffraction data collected at 3 K are refined using {\it FullProf}.\label{fig7}}
\end{figure}

Figure~\ref{fig7} shows the powder neutron diffraction data collected at 3~K and 40~K at Echidna to investigate intensity difference due to magnetic scattering. Figure~\ref{fig7}(a) shows a small shoulder peak for the 3~K data as indicated by the arrow, which corresponds to the (0,2,0) magnetic Bragg reflection. In the inset, the intensity difference between 3~K and 40~K data clearly shows the magnetic Bragg reflections at (1,1,1), (0,2,0), and (3,1,1), and negligible magnetic scattering intensities at (4,0,0) and (6,0,0). This absence of magnetic Bragg intensity at the $(h,0,0)$ reflections is consistent with the single-crystal data discussed previously.
 
\begin{table}
	\caption{\label{irrep}Magnetic representations and their basis vectors for Cu1($x,y,z$), Cu2($-x,-y,z$), Cu3($x+1/4,-y+1/4,z+1/4$), and Cu4($-x+1/4,y+1/4,z+1/4$) (see Fig.~\ref{fig1}).}
	\centering
	\begin{tabular}{c c c c c c c c c c c c c c}
		\hline \hline
		&  & \multicolumn{3}{c}{Cu1} & \multicolumn{3}{c}{Cu2} & \multicolumn{3}{c}{Cu3}& \multicolumn{3}{c}{Cu4}\\
		\hline
		IR & BV & $m_{a}$ & $m_{b}$ & $m_{c}$ &$m_{a}$ & $m_{b}$ & $m_{c}$ &$m_{a}$ & $m_{b}$ & $m_{c}$ &$m_{a}$ & $m_{b}$ & $m_{c}$\\
		$\Gamma_{1}$ & $\psi_{1}$ & 1 & 0 & 0 &    -1 & 0 & 0 &    1 & 0 & 0 &    -1 & 0 & 0\\
		& $\psi_{2}$ & 0 & 1 & 0 &    0 & -1 & 0 &     0 & -1 & 0 &    0 & 1 & 0\\
		& $\psi_{3}$ & 0 & 0 & 1 &    0 & 0 & 1 &     0 & 0 & 1 &    0 & 0 & 1\\
		
		$\Gamma_{2}$ & $\psi_{1}$ & 1 & 0 & 0 &    -1 & 0 & 0 &    -1 & 0 & 0 &    1 & 0 & 0\\
		& $\psi_{2}$ & 0 & 1 & 0 &    0 & -1 & 0 &     0 & 1 & 0 &    0 & -1 & 0\\
		& $\psi_{3}$ & 0 & 0 & 1 &    0 & 0 & 1 &     0 & 0 & -1 &    0 & 0 & -1\\
		
		$\Gamma_{3}$ & $\psi_{1}$ & 1 & 0 & 0 &    1 & 0 & 0 &    1 & 0 & 0 &    1 & 0 & 0\\
		& $\psi_{2}$ & 0 & 1 & 0 &    0 & 1 & 0 &     0 & -1 & 0 &    0 & -1 & 0\\
		& $\psi_{3}$ & 0 & 0 & 1 &    0 & 0 & -1 &     0 & 0 & 1 &    0 & 0 & -1\\
		
		$\Gamma_{4}$ & $\psi_{1}$ & 1 & 0 & 0 &    1 & 0 & 0 &    -1 & 0 & 0 &    -1 & 0 & 0\\
		& $\psi_{2}$ & 0 & 1 & 0 &    0 & 1 & 0 &     0 & 1 & 0 &    0 & 1 & 0\\
		& $\psi_{3}$ & 0 & 0 & 1 &    0 & 0 & -1 &     0 & 0 & -1 &    0 & 0 & 1\\
		
		\hline \hline
	\end{tabular}
\end{table}
\begin{figure}
	\includegraphics[width=8cm]{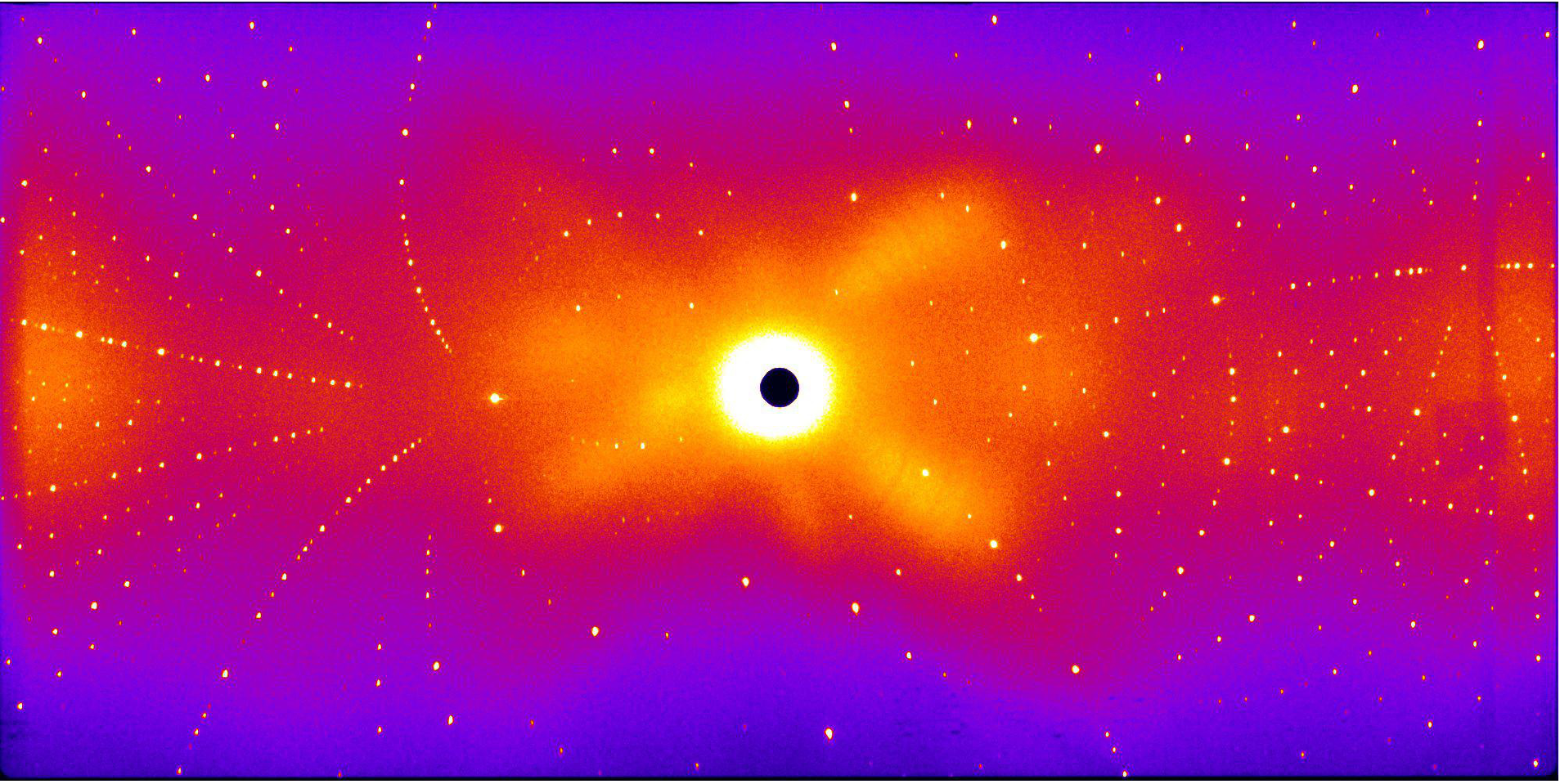}
	\caption{(Color online) False color representation of the single-crystal neutron diffraction measured at 4 K on the Laue diffractometer Koala. The halo around the center of the image is due to scattering from a polycrystalline aluminium crystal holder.\label{fig8}}
\end{figure}

The magnetic structure of $\alpha$-Cu$_{2}$V$_{2}$O$_{7}$ was analyzed by the irreducible representation theory. The calculations were carried out using the software {\sl BasIreps}.\cite{fullprof} The decomposition of the irreducible representations (IRs) for Cu$^{2+}$ ions (16$b$) can be described by
\begin{equation}
\Gamma = 3\Gamma_{1} + 3\Gamma_{2} + 3\Gamma_{3} + 3\Gamma_{4}.
\end{equation}
The basis vectors for each IR are summarized in Table~\ref{irrep}. The coupled intra-chain Cu$^{2+}$ spins are between Cu1--Cu4 along [011], and between Cu2$-$Cu3 along [01$\bar{1}$]. As discussed above, the antiferromagnetic spin component of the Cu$^{2+}$ ions is predominantly along the $a$-axis. Therefore, among the four possible magnetic models we can simply rule out $\Gamma_{2}$ and $\Gamma_{3}$, which according to the Bertaut's notation\cite{Bertaut} give rise to the $A_{x}$ and $F_{x}$ configurations, respectively; both IRs give ferromagnetic spin component along the $a$-axis. On the other hand, $\Gamma_{1}$ and $\Gamma_{4}$, which give rise to the $G_{x}$ and $C_{x}$ configurations, respectively, result in the antiferromagnetic arrangement along the $a$-axis for the spins along the zigzag chains. The canted moments, which are in the $bc$-plane, of the spins in the same chain for $\Gamma_1$ and $\Gamma_4$ are parallel.  Hence the cross product $\vec{S}_i\times\vec{S}_j$ does not have a component along the $a$-axis, making the DM component along the $a$-axis irrelevant as previously stated. Both $\Gamma_1$ and $\Gamma_4$ can give rise the weak ferromagnetism observed in the magnetic susceptibility.  However, $\Gamma_1$ gives a better fit to the powder neutron diffraction data. Figure~\ref{fig7}(b) shows the full pattern refinements for $\Gamma_{1}$ with the spin component along the $a$-axis $m_{x}$ as the only fit parameter while the other two components, {\it i.e.} $m_{y}$ and $m_{z}$, were fixed to zero. The magnetic $R-$factors from the refinements for $\Gamma_{1}$ and $\Gamma_{4}$ yield 0.040 and 0.134, respectively, attesting to the validity of $\Gamma_1$ over $\Gamma_4$. The obtained magnetic moment along the $a$-axis $m_{x}$ is 0.9(2)~$\mu_{B}$ for $\Gamma_{1}$. 
 
\begin{figure}
	\includegraphics[width=8cm]{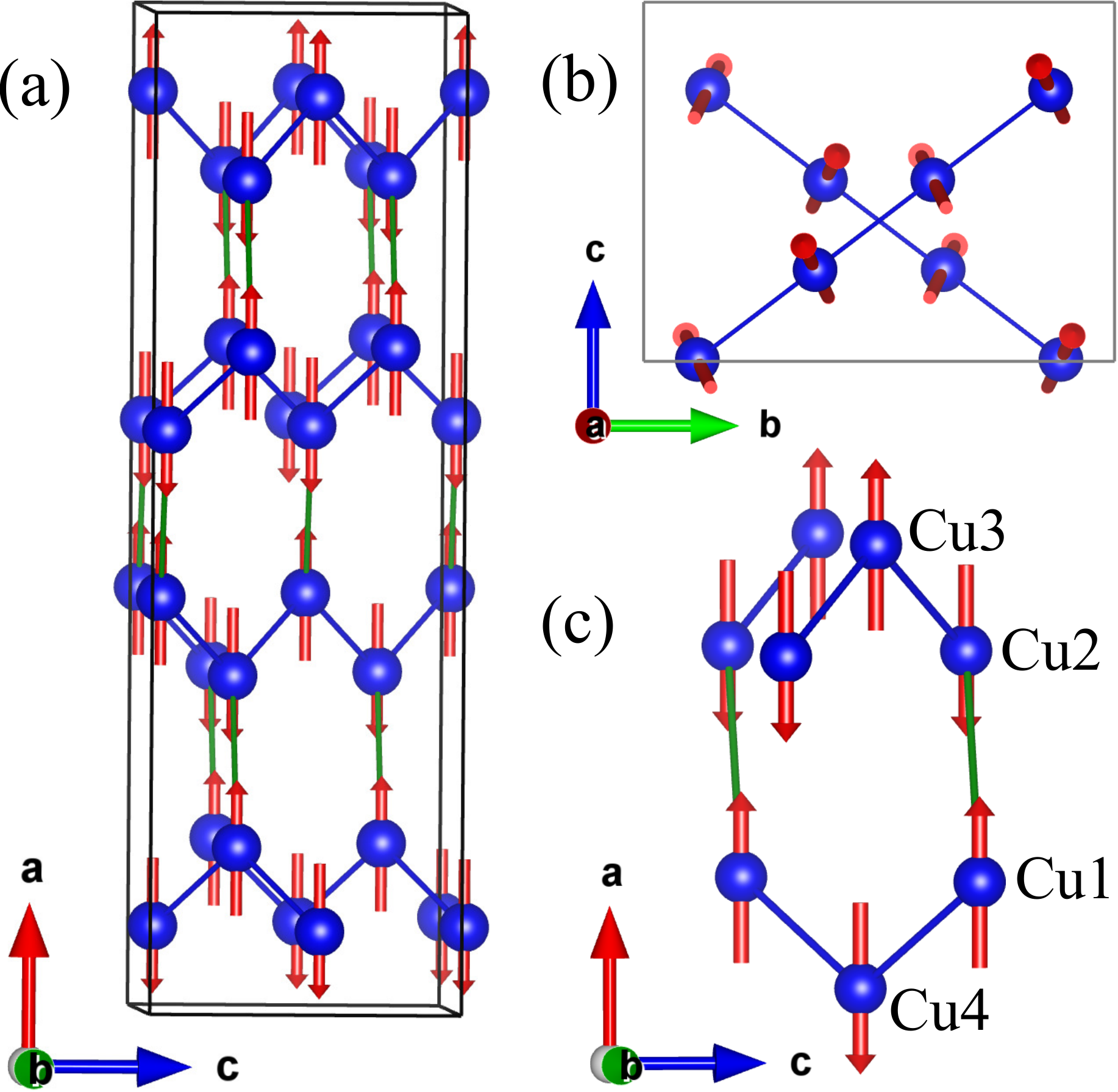}
	\caption{(Color online) (a) The magnetic structure of the $S=1/2$ Cu$^{2+}$ spins in $\alpha$-Cu$_{2}$V$_{2}$O$_{7}$ shows the major spin component along the $a$-axis with the antiferromagnetic arrangement. Blue bonds represent $J_{\text{1}}$ while green bonds $J_{\text{2}}$. (b) The two zigzag chains connected by $J_1$ ($J_2$ not shown) on adjacent $bc$-planes are along [011] and [01$\bar{1}$] and are about 75$^\circ$ with respect to each other. We note that the canted moments in the $bc$-plane are exaggerated for visualization, and the drawn spins do not represent the actual canting, neither in terms of magnitude nor direction. (c) The connectivity of $J_1$ and $J_2$ gives rise to the helical-honeycomb pattern when viewed along the $b$-axis.\label{fig9}}
\end{figure}

To further confirm the magnetic structure and obtain a better estimate of the ordered moment, we performed single-crystal neutron diffraction at the Laue diffractometer, Koala. The diffraction data were collected at 4~K and 50~K to investigate the nuclear and magnetic structure. The Laue diffraction pattern measured at 4~K shows distinct Bragg peaks (Fig.~\ref{fig8}). The structural parameters for the 50 K data were refined against $Fdd2$ space group using {\sl ShelXle} yielding {\sl R$_{1}$}~=~0.081 for 998 reflections with $F_{obs} < 4\sigma(F_{obs}$). The result is in agreement with the single-crystal x-ray diffraction refinement (see Table~\ref{param}). For the 4~K data, where the system becomes magnetically ordered, the magnetic structure refinements were performed using {\it Jana2006}.\cite{jana} The reflections were refined against Shubnikov space group $Fd'd'2$, which is equivalent to the irreducible representation $\Gamma_{1}$ that gives the best fit to the magnetic structure in this system. The spin components along the $b$ and $c$ axes were fixed to zero due to the unresolved spin canting. The ordered moment along the $a$-axis $m_{x}$ = 0.93(9)~$\mu_{B}$ was obtained with $wR$ = 0.051. This value of the ordered moment, which is consistent with the value obtained from powder neutron diffraction, is slightly lower than (but close to) the expected value of one $\mu_{B}$. This discrepancy could be a result from the constrained spin component to only the $a$-axis, discarding the spin canting that is not obtainable from the neutron diffraction data. In addition, quantum fluctuations might also play a role in reducing the ordered moment. The antiferromagnetic spin structure on the helical-honeycomb lattice is depicted in Fig.~\ref{fig9}(a). The spins on the parallel zigzag chains ferromagnetically align as a result of the antiferromagnetic coupling $J_2$ around the helical-honeycomb loop as shown in Fig.~\ref{fig9}(c), allowing the ordering to propagate along the transverse directions and prompting the $3D$ long-range order. Due to the very small spin canting we cannot determine the magnitude and direction of the canted moments in the $bc$-plane from the neutron diffraction data. However, $\Gamma_1$ allows the canted moments of the spins on the same chain to be parallel. For the spins on the different chains that are located on the adjacent $bc$-plane, the $c$-component ($b$-component) is parallel (anti-parallel) as shown in Fig.~\ref{fig9}(b).

\section{Summary}\label{summary}
We have proposed a new spin model to describe magnetic properties of $\alpha$-Cu$_{2}$V$_{2}$O$_{7}$. Combined studies of magnetization, QMC simulations, and neutron diffraction show the helical-honeycomb pattern of spin network connected by two different exchange couplings. Magnetic susceptibility shows a broad peak at $\sim50$~K, which is an evidence of rising short-range spin correlations, followed by an abrupt increase indicative of a phase transition to a magnetically ordered state at $T_{N}$ = 33.4(1) K. The Bethe ansatz calculations for the $S=1/2$ uniform Heisenberg chain fit the $H\perp a$ data very well above $T_N$ but our proposed helical-honeycomb model shows substantial improvement of the fit for $J_{1}:J_{2} = 1:0.45$ with $J_1=5.79(1)$~meV and for $J_{1}:J_{2} = 0.65:1$ with $J_2=6.31(1)$~meV. Therefore, we conclude that the helical-honeycomb model describes the underlying spin network of $\alpha$-Cu$_{2}$V$_{2}$O$_{7}$ more accurately than the previously held spin-chain model.

The anisotropy below $T_N$ suggests that the majority of the spin component is along the crystallographic $a$-axis, which is confirmed by neutron diffraction experiments. The weak ferromagnetism is a result of spin canting within the $bc$-plane due to the DM interactions. Magnetization measurements with $H\perp a$ show the spontaneous magnetization from which the canting angle $\eta$ of 4.0$^\circ$ and the in-plane DM parameter $\left|D_p\right|\simeq0.14J_1$ are obtained.

The analysis of the neutron diffraction data shows that the $S=1/2$ Cu$^{2+}$ spins antiferromagnetically align along the helical-honeycomb loops with the ordered moment of $0.93(9)~\mu_B$ predominantly along the crystallographic $a$-axis. The spin network of two comparable exchange couplings forming the helical-honeycomb lattice and the DM interactions lead to the long-range magnetic ordering below $T_N$. However, due to the complex exchange pathways and the presence of weak frustration, the exact value of both exchange interactions could deviate from our analysis. Further theoretical analyses based on first-principle calculations and studies of spin dynamics by means of inelastic neutron scattering are therefore required in order to better determine the exchange interactions and confirm the helical-honeycomb spin model.  With the current availability of large single crystals, an inelastic neutron scattering study is possible and will reveal a complete picture of the relevant microscopic Hamiltonian parameters, as well as influences of the low spin-coordination number and quantum fluctuations on spin dynamics.

{\it Note added in proof} After the submission of this manuscript, we became aware of the work by Sannigrahi {\it et al.}\cite{sannigrahi}, in which they performed density functional theory to calculate the relevant exchange interactions in $\alpha$-Cu$_2$V$_2$O$_7$.  However, the QMC calculations of the susceptibility show that their lattice model with the dominant third-nearest-neighbor antiferromagnetic interactions ($J_3=13.61$meV) gives a broad peak at a higher temperature than $\sim50$ K, inconsistent with the experimental data.
\vspace{10mm}

\begin{acknowledgments}
Work at Mahidol University was in part supported by the Thailand Research Fund Grant Number MRG5580022 and the Thailand Center of Excellence in Physics. Work at IMRAM was partly supported by Japan Society for the Promotion of Science KAKENHI Grant Number 24224009. The authors, G. G. and K. M., would like to thank I. M. Tang and T. Osotchan for the use of their laboratory equipment and M. A. Allen for useful discussions. The identification of any commercial product or trade name does not imply endorsement or recommendation by the National Institute of Standards and Technology.
\end{acknowledgments}

\bibliography{reference}
\end{document}